# Nonlinear Mode Coupling and Internal Resonances in $MoS_2$ Nanoelectromechanical System

C. Samanta, G.P.R. Yasasvi and A. K. Naik

*Centre for Nano Science and Engineering, Indian Institute of Science, Bangalore, 560012, India*

**Abstract:**

**Atomically thin two dimensional (2D) layered materials have emerged as a new class of material for nanoelectromechanical systems (NEMS) due to their extraordinary mechanical properties and ultralow mass density. Among them, graphene has been the material of choice for nanomechanical resonator. However, recent interest in 2D chalcogenide compounds has also spurred research in using materials such as $MoS_2$ for NEMS applications. As the dimensions of devices fabricated using these materials shrink down to atomically thin membrane, strain and nonlinear effects have become important. A clear understanding of nonlinear effects and the ability to manipulate them is essential for next generation sensors. Here we report on all electrical actuation and detection of few layers $MoS_2$ resonator. The ability to electrically detect multiple modes and actuate the modes deep into nonlinear regime enables us to probe the nonlinear coupling between various vibrational modes. The modal coupling in our device is strong enough to detect three distinct internal resonances.**

Sensors made of nanoelectromechanical devices are now capable of measuring mass of individual protein molecules[1,2] and have resolution down to atomic mass unit[3]. These devices are also fantastic tool to probe validity of continuum mechanics at atomic thickness[4,5], study nonlinear dynamics[6,7] and gain deeper insight into quantum mechanics[8–10]. Among the 2D materials employed for fabrication of these devices, graphene has drawn the most attention and has been extensively studied[11–17]. However, recent theoretical models indicate the possibility of lower dissipation in other 2D materials such as $MoS_2$[18]. Furthermore, ultralow mass density of 3.3fg/µm$^2$, high elastic modulus ($E_y \approx 0.3$ TPa)[19,20] and an exceptional strain limit of 10-12%[20] make it an attractive alternative to graphene for NEMS applications. The strongly coupled mechanical and electro-optical properties of $MoS_2$[21–23] will also have implications for optomechanics and valleytronics[24,25]. For these reasons there has been effort towards characterizing the properties of nanoelectromechanical devices made of $MoS_2$[26–28] . The experimental research reported till date has relied on optical detection primarily due to the exquisite sensitivity to motion as well as ability to detect multiple modes of the device. However, characterization of these devices deep into the nonlinear regime has proven difficult. Furthermore, integration of optical detection scheme for application is cumbersome and an all electrical actuation and detection scheme would be desirable. In this report, we demonstrate all electrical actuation and detection of few layers $MoS_2$ resonator. Similar to optical detection schemes, we are able to observe multiple vibrational modes of the device and unlike previous reported work on 2D materials we are able to drive these devices deep into nonlinear regime to observe nonlinear coupling. Due to strong nonlinear coupling between different vibrational modes, we also observe multiple internal resonances[29–32]. Although there has been one report of internal resonances in microelectromechanical system (MEMS)[32], there is no reported evidence of internal resonances in resonators made of thin atomic membrane. The large strain observed in these ultra-thin devices and the associated strong modal coupling observed in this work has major implication for sensing, novel low noise oscillators[32] and nonlinear dynamics[33].

The device used in this study is a few layer (bi-layer or tri-layer based on the contrast) $MoS_2$ resonator. The fabrication procedure is similar to other published reports of ultrathin resonators[12,13] (See SI). The highly doped silicon substrate is utilized as the gate for the actuation of the device. The gate capacitor has approximately 150nm of air gap and about 140nm of silicon oxide. Figures 1a and 1b show optical image and electron micrograph of the device respectively. Electrical characterization of the suspended device yields a mobility of $9cm^2/V\text{-}s$ (See figure S1).

In this work, we utilize three different transduction schemes, viz., 1) 1ω mixed down technique[12,34], 2) 2ω mixed down technique[34,35] and 3) Frequency Modulation (FM) technique[36]. All the three schemes are able to detect multiple vibrational modes of this ultrathin $MoS_2$ nanoresonator and show a fundamental resonance frequency of about 41.8MHz and quality factor of about 600 (figure1c). The measurements reported in this paper are performed at room temperature and vacuum levels below $10^{-7}$ *Torr*.

Figure 2 shows the various resonant modes of the device detected using the three different transduction schemes. In addition to the mechanical resonance, we observe a number of electrical background peaks in 1ω and 2ω mixed down detection schemes. The peaks are identified as mechanical if nonlinear response is observed. For peaks below 100MHz, we are able to actuate the resonator deep into nonlinear regime in all the actuation schemes mentioned above. For peaks above 100MHz, none of the actuation schemes is strong enough to drive the modes non-linear; we rely on the ability of FM technique to pick only the mechanical resonances[36].

Table 1 gives the list of all the mechanical resonances that are observed with the three techniques. In this work we are unable to associate the higher frequencies with mode shapes and as such use the nomenclature 1st mode for the first observed mechanical resonance (41MHz), 2nd mode for 70MHz resonance and so on. Atomically thin suspended membrane fabricated using the so called - scotch tape method typically have large strain[11–13]. The strain affects the resonant frequency and nonlinear coefficients of the device and thus the nonlinear coupling between different modes of the device. To estimate the intrinsic strain of the device, we measured the frequency of 1st mode of the device as a function of back gate voltage and calculated the intrinsic strain and mass loading on the device[17] (see figure S5). Based on the fitting of the experimental data the strain is estimated to be approximately $10^{-2}$ (assuming the device to be bilayer) at room temperature.

Unlike the electrical measurements of graphene resonators reported to date, the ability to observe multiple modes and high strain makes it attractive to study nonlinear coupling between various vibrational modes. These devices can be driven into nonlinear regime by relatively modest electrostatic forces due to their atomically thin nature. Because of the presence of the electrostatic gate, the nonlinear driven resonant mode can be described by asymmetric Duffing oscillator equation given by[29]

$$\ddot{x} + 2\zeta\dot{x} + \omega_{0n}^2 x + \alpha_2 x^2 + \alpha_3 x^3 = \frac{F}{m}\cos(\omega t) \qquad (1)$$

Where, $\zeta$ is the damping ratio, $\omega_{0n}$ is the resonant frequency of the $n^{th}$ mode, $\alpha_2$ and $\alpha_3$ are the quadratic and cubic nonlinearity coefficients and *F* is the drive force.

Figure 3a shows the response of the $1^{st}$ mode of device as the drive amplitude is increased. The initial Lorentzian shape is quickly driven to nonlinear regime with critical amplitude of about *7nm* at dc gate voltage ($V_g^{DC}$) of 15V (See figure S6). Beyond the linear regime, the device shows the hardening nonlinear response. However, for extremely large drives corresponding to about 5-22 times the force required at critical amplitude, we observe signatures of nonlinear mode coupling for two distinct actuation drive ranges. For all ac drive voltages ($V_g^{AC}$) between 0.3V to 1.25V, the downward jump during the forward frequency sweep remains fixed at 42.032MHz (see figure 3b and 3d). For drive level above 1.25V we observe two features, viz., i) a dip appears at 42.032MHz (red arrow in figure 3c and SI figure S7b) and ii) the downward jump is now fixed at 43.12MHz. Figure 3d shows the peak amplitude and position of the frequency jump as a function of the drive voltages. The two distinct plateaus observed both in the peak amplitude as well as the jump frequency point (upper to lower branch) correspond to two distinct internal resonances. Internal resonances are likely to occur in systems with large nonlinear mode couplings leading to spillover of the energy pumped into one mode to other vibrational modes[29]. The plateau width spanning approximately one decade of ac drive indicates an internal resonance that is extremely stable against drive amplitude fluctuations. In presence of coupling between two vibrational modes, the equation for the nonlinear driven mode can be written as

$$\ddot{x}_1 + 2\zeta\dot{x}_1 + \omega_{01}^2 x_1 = \frac{F}{m}\cos(\omega t) + \beta(x_1, x_2) - \alpha_2 x_1^2 - \alpha_3 x_1^3 \qquad (2)$$

Where, $\beta$ is the coupling constant that is a function of displacement of the two modes $x_1$ and $x_2$. The nonlinear coefficient terms that have been rearranged on the right hand side have components at $2\omega$ and $3\omega$. If the higher mode satisfies the condition $\omega_{02} \approx 2\omega_{01}\ or\ \omega_{02} \approx 3\omega_{01}$ , these terms can provide the energy to drive the higher modes leading to internal resonance[29].

To verify the coupling between different vibrational modes we actuate the first mode at a constant drive voltage. The drive frequency of the first mode is then varied in steps while driving and monitoring the frequency response of the 5th mode. The drive amplitude of the 5th mode was maintained to ensure moderate signal to noise levels for this mode but still well below its nonlinear limit. Figures 4a-c show the frequency response of the 1st mode for 3 different drive forces. Figure 4a in linear regime, 4b in nonlinear regime corresponding to first plateau and 4c in nonlinear regime corresponding to second plateau. Figures 4d-f show the frequency shift observed in mode 5 for the corresponding drives of mode 1 shown in figure 4a-c. In all cases, the resonant frequency shift of mode 5 closely follows the amplitude of the 1st mode indicating the amplitude dependent tension as the source of coupling[37,38].

The observed strong coupling and the associated plateaus observed are characteristics of internal resonances. The 1:2 internal resonances are predicted to occur in systems with quadratic nonlinearity and with modes such that one vibrational mode has resonant frequency that is close to twice the resonant frequency of another mode[29]. For our device $\omega_{04,05} \approx 2\,\omega_{01}$. The 1st mode goes into internal resonance with these two vibrational modes at two different drive levels leading to two plateaus shown in figure 3d. To check the possibility of mode 1 coupling to other vibrational modes, we performed similar measurements on mode 2, 3 and 6 and found the coupling with mode 1 to be much weaker (see figure S8).

We also observe strong nonlinear coupling between mode 4 and mode 5 resulting into 1:1 type of internal resonance. In nonlinear Duffing systems with cubic nonlinearities and modes with frequencies close to each other, it is possible to observe 1:1 internal resonances[31]. Figure 5a shows the response of mode 5 as the drive force is progressively increased. The initially linear Lorentzian response slowly transforms into resonance curve with two peaks. This splitting of the curve is a classic example of nonlinear mode coupling resulting into 1:1 internal resonance[31].

It is quite possible that the three internal resonances (and possibly more) in turn are coupled to other vibrational modes leading to complex transfer of energy between them. Detailed information about the various mode shapes, their coupling mechanisms and the ability to experimentally measure thermomechanical noise are essential to explicitly observe and understand the complex nature of the transfer of energy during the internal resonances.

Conclusion:

In conclusion, we demonstrate all electrical actuation and detection of atomically thin $MoS_2$ nanoelectromechanical resonator. Unlike previous reports with electrical measurements of such resonators, we are able to identify more than 10 mechanical modes. This is especially useful in resonators with length smaller than a few micrometers which are difficult to transduce using optical method. Due to the large strain observed in these devices, there is strong nonlinear coupling between various vibrational modes of the devices. We exploit this strong coupling to demonstrate, for the first time in atomically thin resonator, two different types of internal resonances.

This ability to strongly couple different modes has implications for applications such as high stability oscillator[32] and sensors in very high and ultra high frequency range. The strong coupling and the ability to manipulate the coupling by changing strain also makes these devices attractive tool to study nonlinear dynamics and quantum mechanics.

**Acknowledgements**:

We acknowledge financial support in the form of start-up grant from the Indian Institute of Science, Bengaluru. We gratefully acknowledge the usage of National Nanofabrication Facility (NNfC) and Micro and Nano Characterization Facility (MNCF) at CeNSE, IISc, Bengaluru. C.S acknowledges UGC for financial support. The authors thank Villanueva Luis Guillermo and Mandar Deshmukh for helpful discussion.

**Author Contribution**:

C.S fabricated the device, performed major portions of the experiment and analysis. Y.G.P.R. helped in the 2$\omega$ and FM measurements. A.K.N. provided the overall guidance for the project. All authors discussed the data and contributed to writing the manuscript.

**Competing financial interests**:

Authors declare no competing financial interests.

**Figures:**

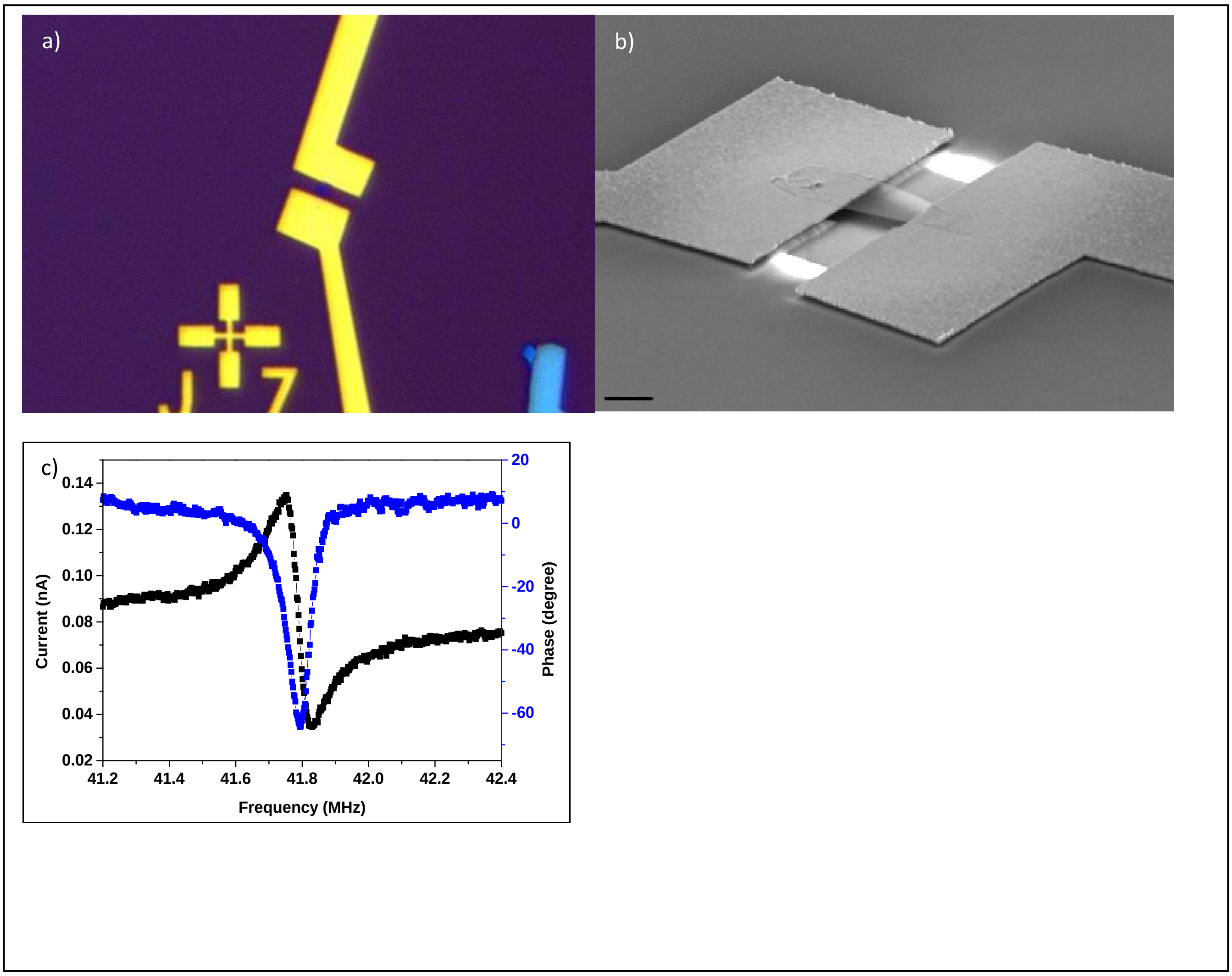

**Figure 1**: **$MoS_2$ nanoresonator** a) and b) show optical image and scanning electron micrograph of the $MoS_2$ resonator respectively. Scale bar is 1µm. c) Magnitude and phase response of the first mode of the $MoS_2$ resonator using 1ω method with mixed down frequency of 1005Hz with $V_s^{AC}$ =71mV , $V_g^{AC}$=58 mV and $V_g^{DC}$= 6V

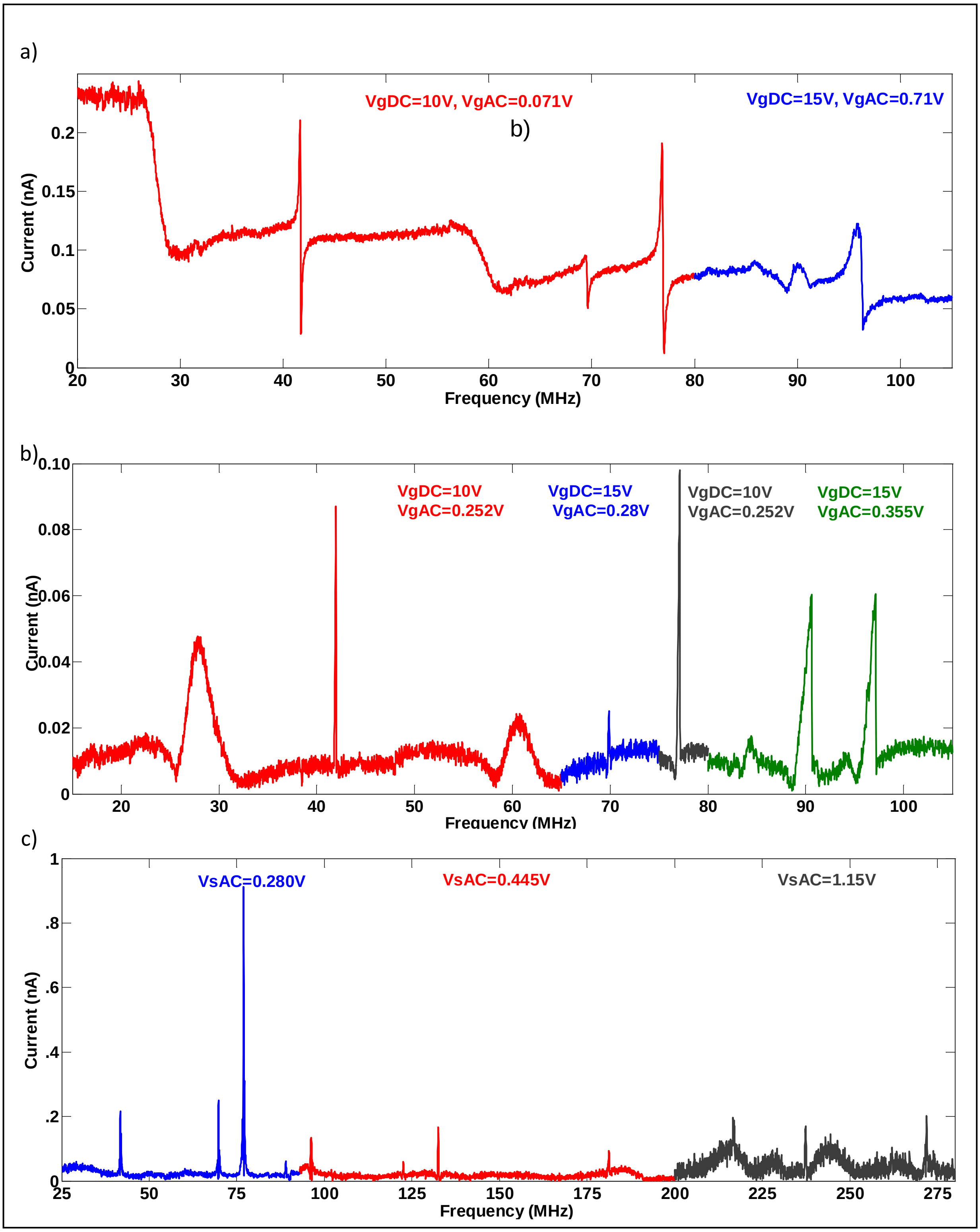

**Figure 2**: **Observed vibrational modes of the $MoS_2$ nanoresonator:** The vibrational modes observed using a) 1ω measurement technique with $V_s^{AC}$ of 71mV, b) 2ω measurement scheme with $V_s^{AC}$ of 71mV and c) FM modulation measurement technique with $V_g^{DC}$=12V. In both 1ω and 2ω methods mixed down frequency of 1005Hz is used. In FM measurement frequency deviation of 200kHz and modulation frequency of 1005Hz are used. The broad peaks in the measurements are due to electrical background.

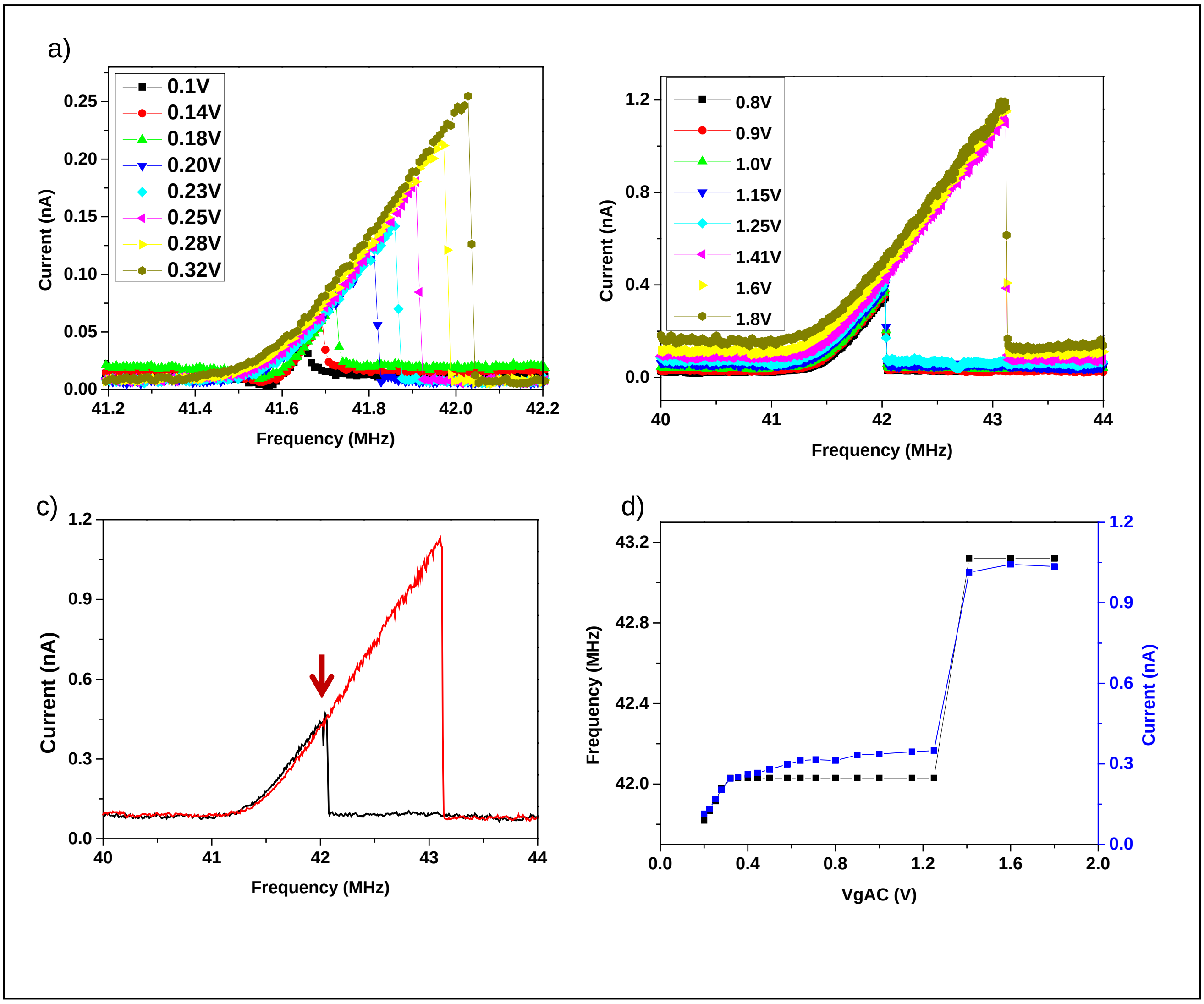

**Figure 3**: **Nonlinear response of the device**: a) Typical forward sweep frequency response of the device at moderate drives shows the nonlinear behavior. b) Response of device during the two internal resonances. Pinning of the peak frequency at 42.032MHz and 43.12MHz for a wide range of drive levels is observed. c) Response of the first mode during forward (red) and backward (black) frequency sweep. The peak frequency 43.12MHz in forward sweep corresponds to second internal resonance and a dip in backward sweep at 42.032MHz corresponds to the first internal resonance. d) Each internal resonance is characterized by pinning of the peak frequency and minimal change in peak amplitude for a large range of AC drive forces. The measurements were performed using 2ω method with $V_g^{DC}$=15V and $V_s^{AC}$=71mV.

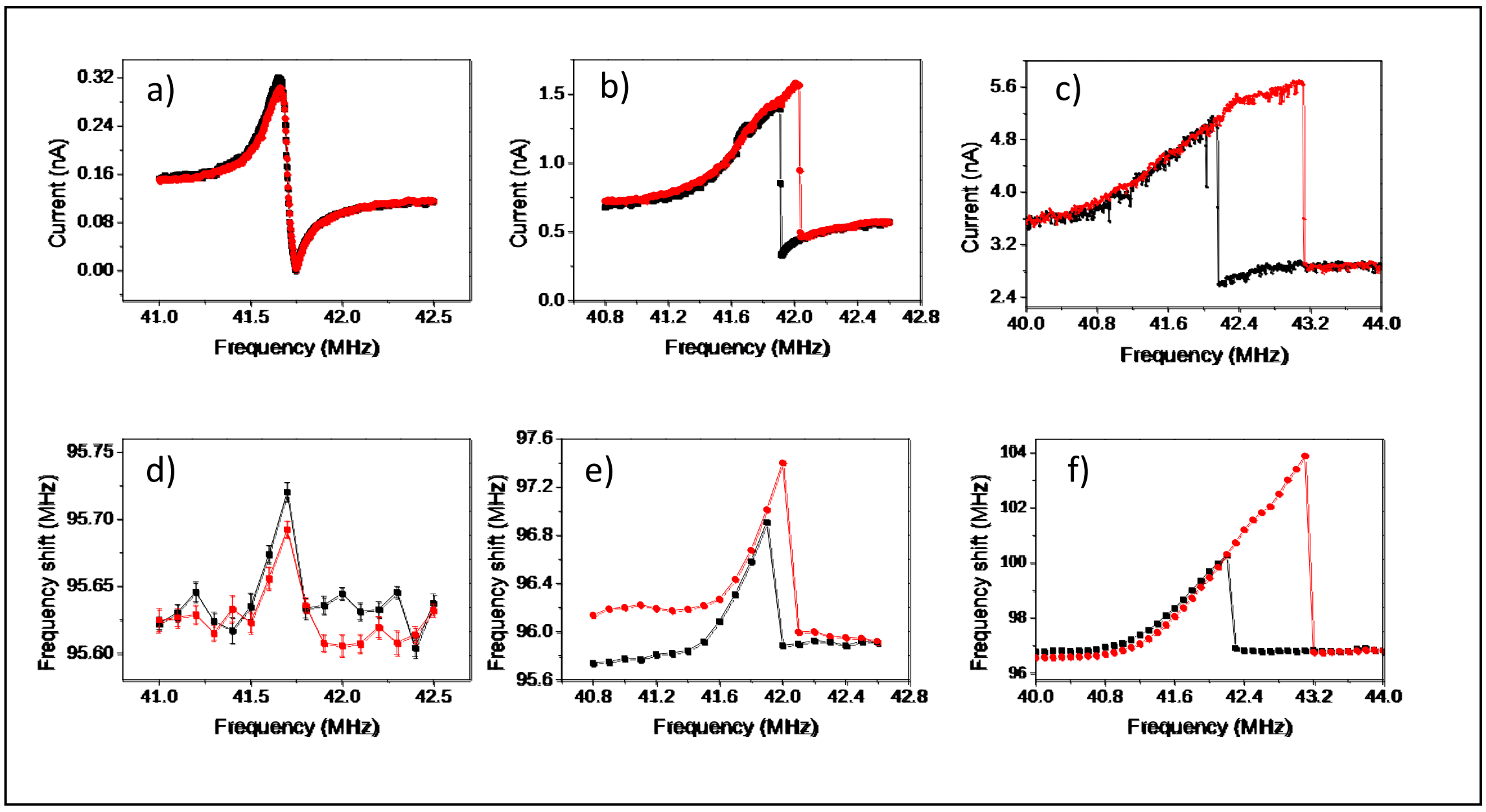

**Figure 4**: **Mode coupling.** Response of the mode 1 for forward (red) and backward (black) frequency sweeps for drive levels in linear regime (a), for drive levels corresponding to 1$^{st}$ internal resonance (b) and for drive levels corresponding to 2$^{nd}$ internal resonance (c).(d)-(f) show the shift in resonant frequency of mode 5 for the three drives shown in (a)-(c). 1ω method is used in all the cases.

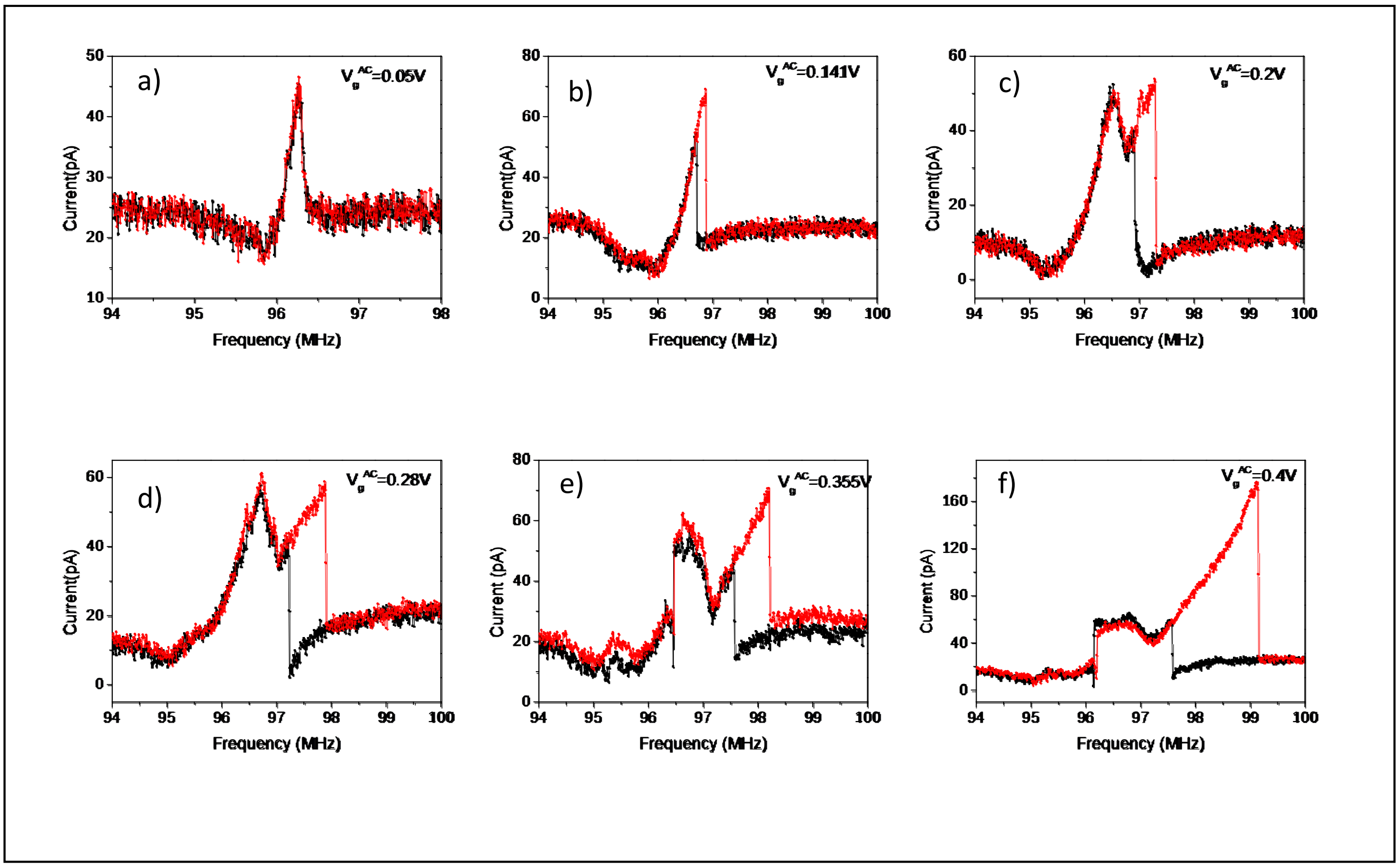

**Figure 5**: **1:1 internal resonance between mode 4 and mode 5**. Electromechanical response of mode 5 for forward and backward frequency sweeps for different drive levels. As the drive level for mode 5 is increased a single peak splits into two due to strong mode to mode coupling. The measurements were performed using 2ω method with $V_g^{DC}$=12V and $V_s^{AC}$=0.07V.

| Mode Number | | | 1 | 2 | 3 | 4 | 5 | 6 | 7 | 8 | 9 | 10 | 11 |
|---|---|---|---|---|---|---|---|---|---|---|---|---|---|
| Nominal $f_{res}$(MHz) | | | 41 | 70 | 77 | 89 | 96 | 122 | 131 | 181 | 218 | 237 | 272 |
| Transduction schemes | 1 ω | $f_{res}$ | 41.8 | 69.8 | 77.0 | 88.9 | 95.9 | NA | 132.4 | NA | NA | NA | NA |
| | | Q | 680 | 506 | 550 | 320 | 340 | | 440 | | | | |
| | 2 ω | $f_{res}$ | 41.8 | 69.8 | 76.9 | 88.9 | 96.0 | NA | NA | NA | NA | NA | NA |
| | | Q | 690 | 620 | 537 | 476 | 403 | | | | | | |
| | FM | $f_{res}$ | 41.8 | 69.7 | 76.9 | 88.8 | 96.2 | 122.6 | 132.5 | 181.5 | 216.8 | 237.3 | 272 |
| | | Q | 533 | 474 | 511 | 255 | 270 | 265 | 316 | 350 | 243 | 280 | 473 |

**Table 1:** Modes observed using different transduction schemes

# Supplemental Material

# Nonlinear mode coupling and internal resonances in $MoS_2$ Nanoelectromechanical system

*C. Samanta, G.P.R. Yasasvi and A. K. Naik*

*Centre for Nano Science and Engineering, Indian Institute of Science, Bangalore, 560012, India*

**Device Fabrication:**

Bulk $MoS_2$ was exfoliated on 285nm thick $SiO_2$ on a highly p-doped Si substrate using standard scotch-tape technique. $MoS_2$ flakes were identified under optical microscope. Contact pads were patterned by electron beam lithography followed by thermal evaporation to deposit Cr/Au (5nm/70nm). A second layer of lithography was done to open a small window for selective etching of $SiO_2$ under the $MoS_2$ flake. Buffered Oxide Etchant (BOE) was used to etch out approximately 150 nm of $SiO_2$ under the flake. Subsequently critical point drying (CPD) was carried out to ensure that the suspended flake does not collapse due to surface tension. The device is slightly asymmetric with length of about 1.6μm and approximate width of 1.3μm.

**Electrical characterization:**

We perform electrical characterization of the suspended $MoS_2$ device using two Keithley source meters. A dc voltage $V_{ds}$ is applied to the source contact pad keeping the drain contact pad grounded while another dc voltage $V_g^{DC}$ is applied to highly doped Si substrate as global back gate voltage. Figure S1a shows the output characteristic ($I_{ds}$-$V_{ds}$) at different gate voltages and Figure S1b shows transfer characteristic ($I_{ds}$-$V_g$) of the device. All the electrical measurements are carried out at room temperature and less than $10^{-7}$ Torr vacuum. Forward & backward sweep data are taken for all the measurements and there is no sign of hysteresis. The symmetric nature of $I_{ds}$-$V_{ds}$ indicates that the contacts are ohmic in our operating range of the $V_{ds}$ (-1V to +1V).

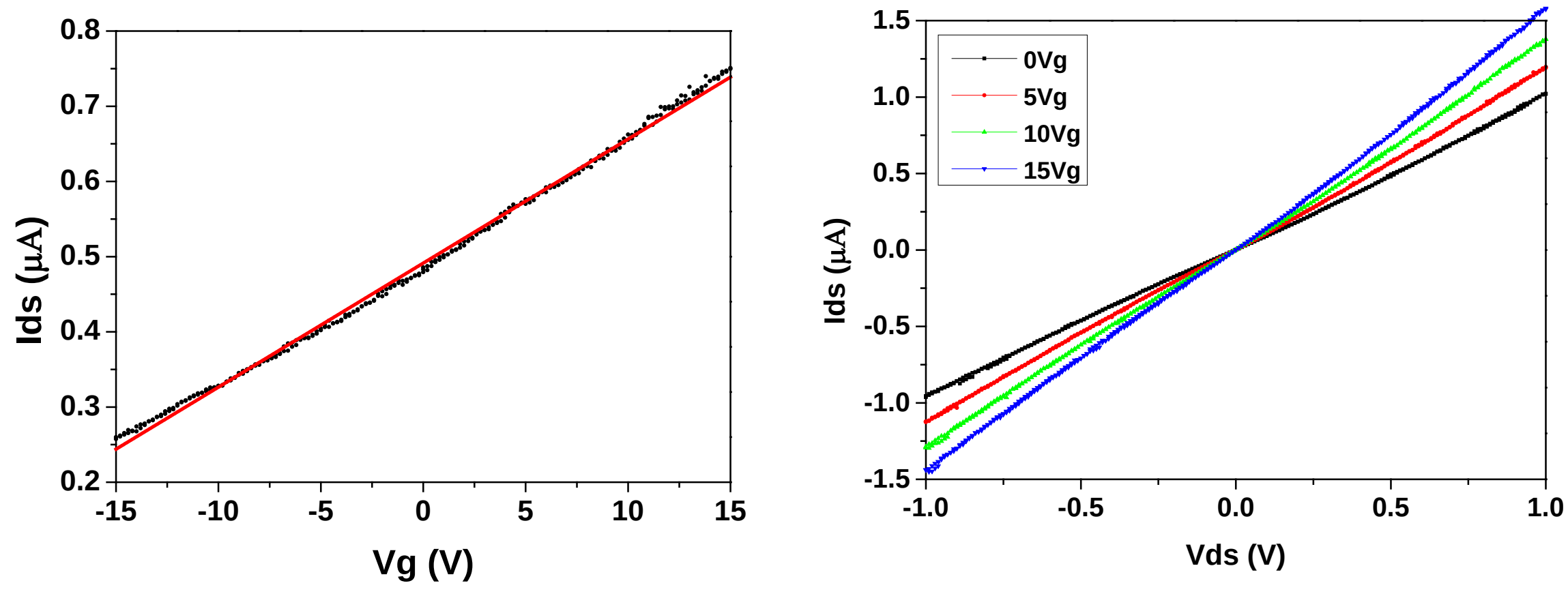

**Figure S1**: *IV characteristics of the suspended $MoS_2$ device. a) Variation of drain current with gate voltage is used to calculate the mobility. b) $I_{ds}$-$V_{ds}$ curves for different gate voltages.*

Field-effect mobility ($\mu_{FE}$) was extracted using the equation

a) $$\mu_{FE} = \frac{L}{WCV_{ds}} \frac{dI_{ds}}{dV_g} \quad (1)$$ b)

where *L* is channel length, *W* is channel width and *C* is the capacitance per unit area between channel and back gate[1]. The mobility of the device is estimated to be 9cm$^2$/V-s.

**Electro-Mechanical characterization:**

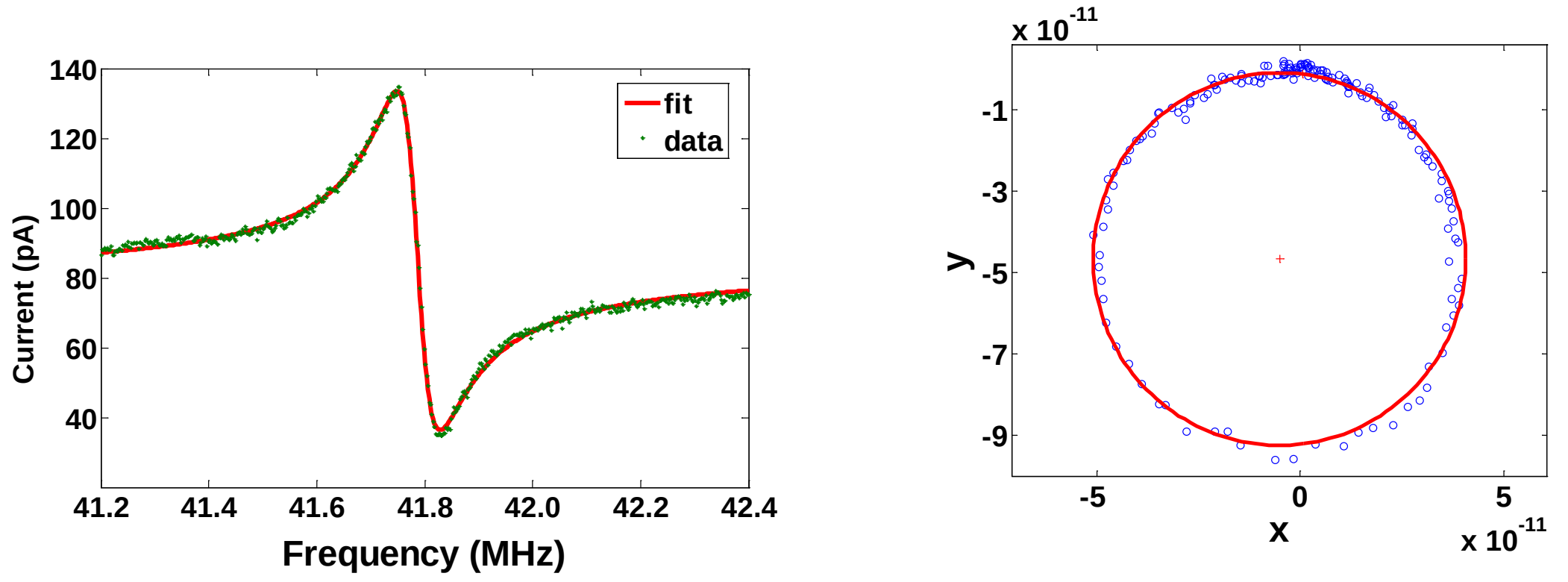

**Figure S2**: *The response of the fundamental mode. Red solid line is the fit to the experimental data.*

In this work, we rely on three different transduction mechanisms viz 1) 1ω mixed down technique[2,3], 2) 2ω mixed down technique[3,4] and 3) Frequency Modulation(FM) technique[5]. As shown in table 1 in the main text, the three transduction schemes are reliably able to detect most mechanical resonances below 100MHz, although with different signal to noise ratio. Figure S2 shows the response of the fundamental mode obtained using 1ω mixed down method. The quality factor and the resonance frequency of the mode are obtained by fitting the experimental data. Figure S3 shows the measurement of the same mode using FM technique. The frequency deviation in this measurement was adjusted to obtain reasonable signal to noise while avoiding measurement related broadening associated with this technique[5]. Figure S4 shows the in-phase and out of phase components of the response of the device. Matlab fit to the in-phase response[5] is used to extract the resonant frequency and the quality factor. In this work the choice of transduction scheme for various measurements is primarily determined by the signal to noise requirements as well as the ability to drive the mode deep into nonlinear regime.

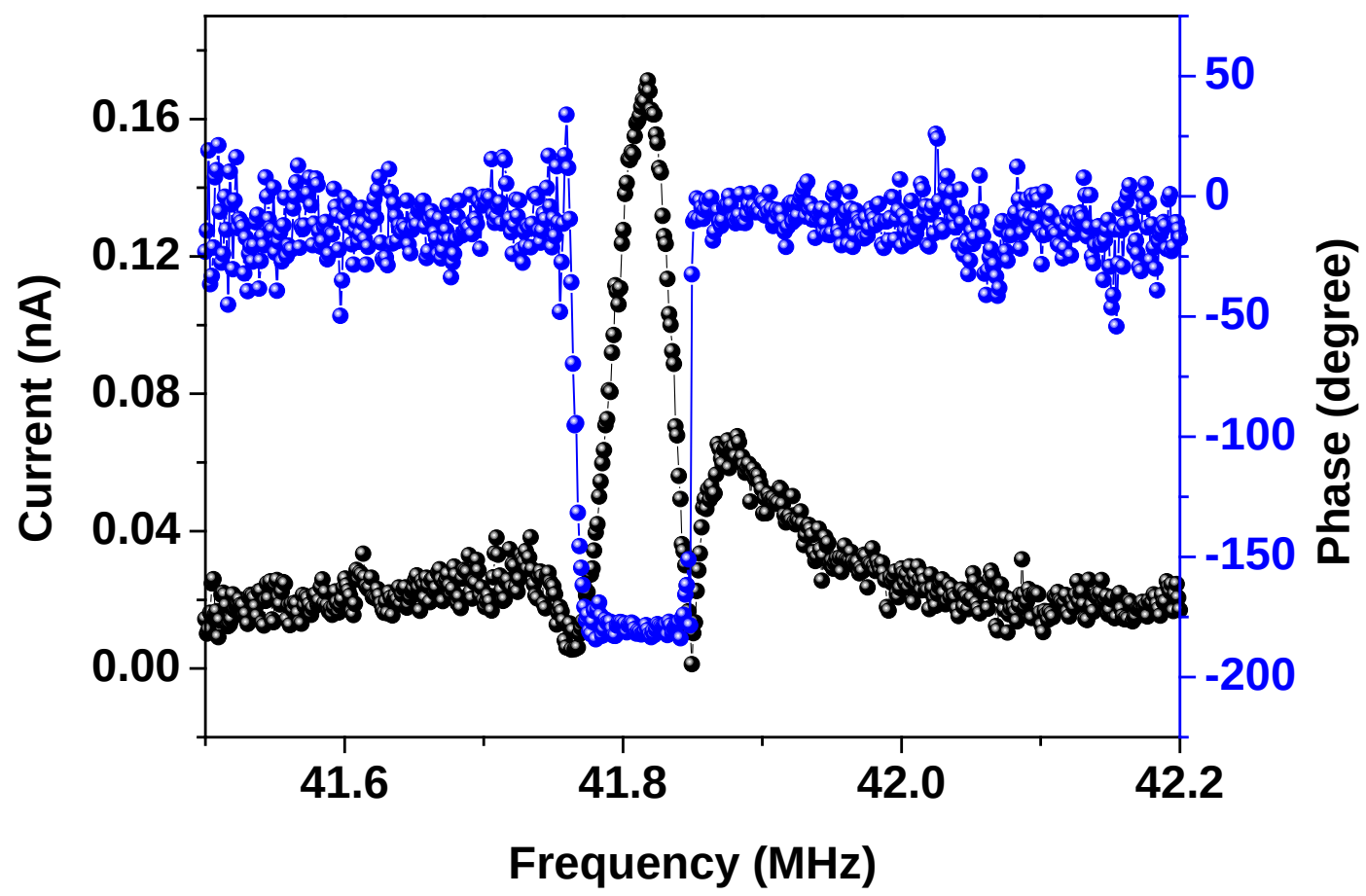

**Figure S3**: *Amplitude and phase response of the fundamental mode using FM technique. The two side lobes in the amplitude and 180 degree phase shift in phase are clearly visible.*

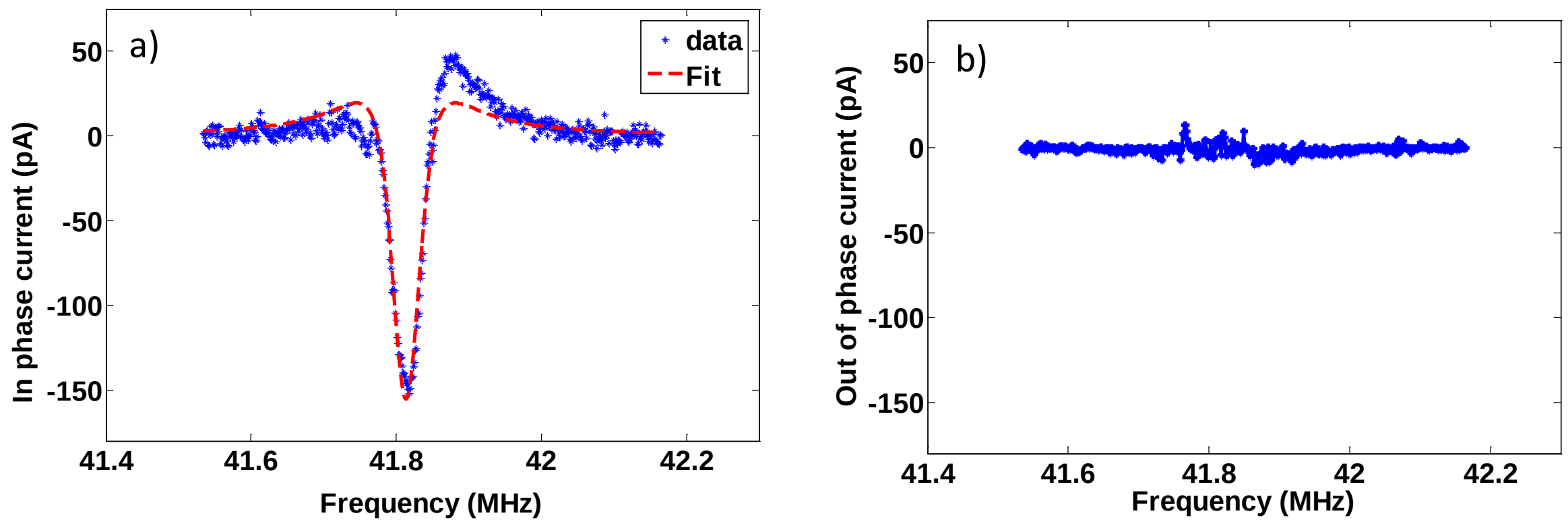

**Figure S4**: *a) and b) show in phase and out of phase response of the device observed by FM technique respectively.*

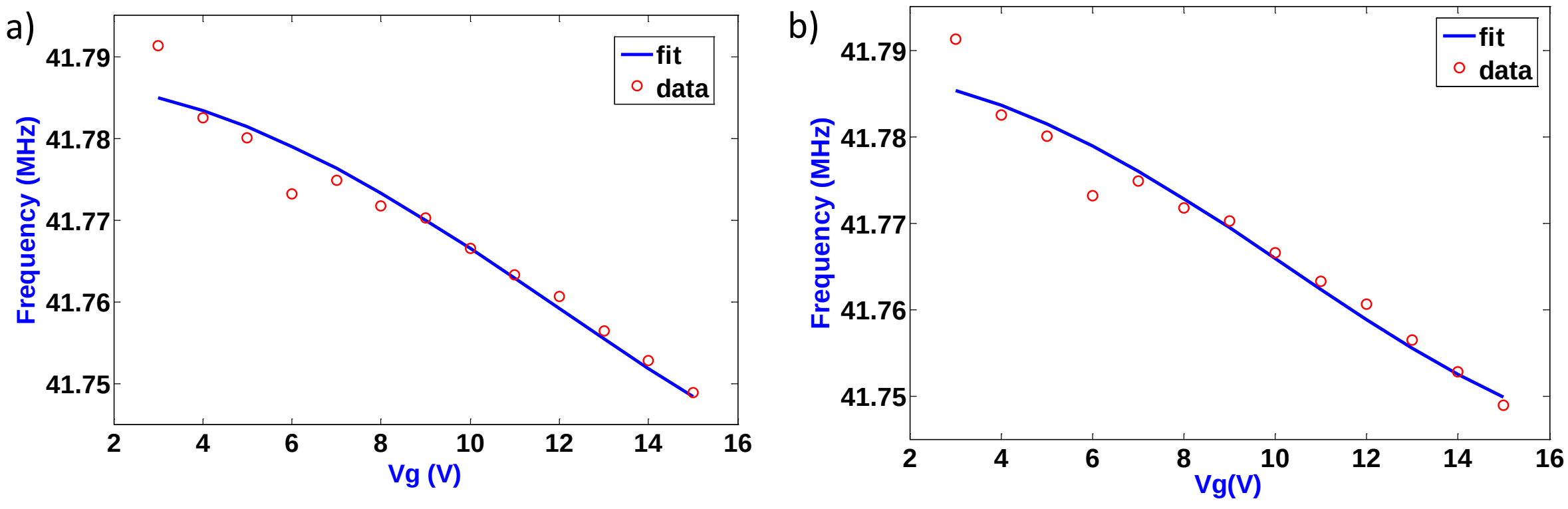

**Figure S5**: *Frequency response of the fundamental as a function of back gate dc voltage. Experimental data is shown as red filled circles and the line is the fit. Fitting yields slightly different value of strain depending on if the layer is bi-layer (a) or tri-layer (b).*

**Strain calculations:**

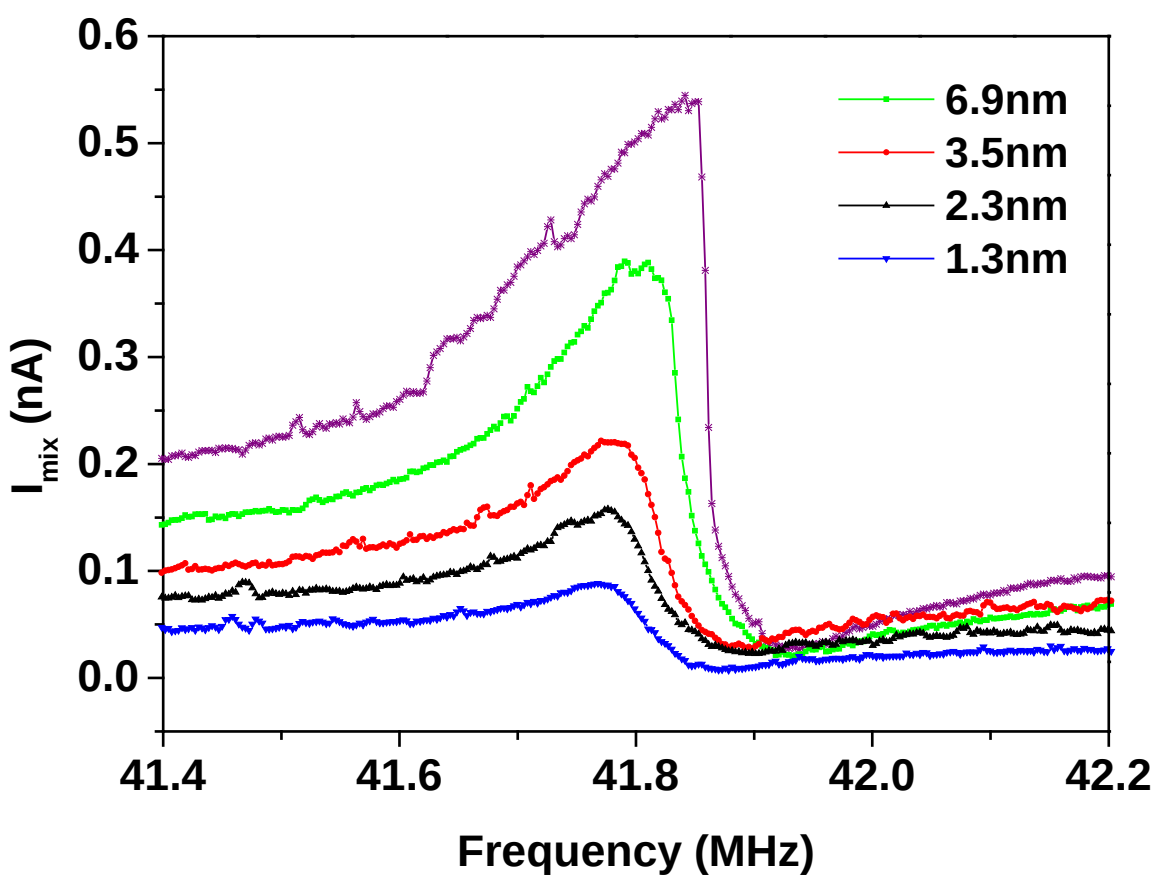

**Figure S6**: *The amplitude response of the device using 1$\omega$ measurement technique. For drive level ($V_g^{AC}$=115mV) corresponding to purple curve we see a hysteric response in forward and reverse frequency sweeps. In these measurements we bias the device with $V_s^{AC}$=71mV and $V_g^{DC}$=15V.*

Based on the optical contrast, our device is either bilayer or tri-layer. The intrinsic strain in the device is determined by fitting the resonant frequency change with DC gate voltage (see figure S5). Assuming it to be a bilayer we estimate the strain to be about 0.9% with mass loading of about 30. The strain is estimated to be about 0.7% with mass loading of about 20 if the device is assumed to be tri-layer.

**Critical amplitude calculation:**

The displacement of the resonator in the linear regime is calculated using the procedure described elsewhere[2]. We estimate the critical amplitude of the device to be close to 7nm. This relatively large critical amplitude could be the result of large strain as well as the interplay of both the cubic and quadratic nonlinearities in the system[6].

**Internal resonance using 1ω method:**

Response of the fundamental mode in the strongly nonlinear regime can also be studied using the 1ω method. Figure S7 shows the response of the device in the nonlinear regime. The pinning of the frequency at 42.032MHz can be clearly seen. We also observe a small dip at 42.032MHz in the forward and backward frequency sweep when the drive levels are larger. This dip is typically larger during the backward sweep. In this method we observe increase in peak amplitude with ac drive level ($V_g^{AC}$) during the first internal resonance. We believe the change in amplitude observed during internal resonance (figure S7c) is primarily due to the change in electrical background. The electrical background is larger in 1ω method compared to the 2ω method.

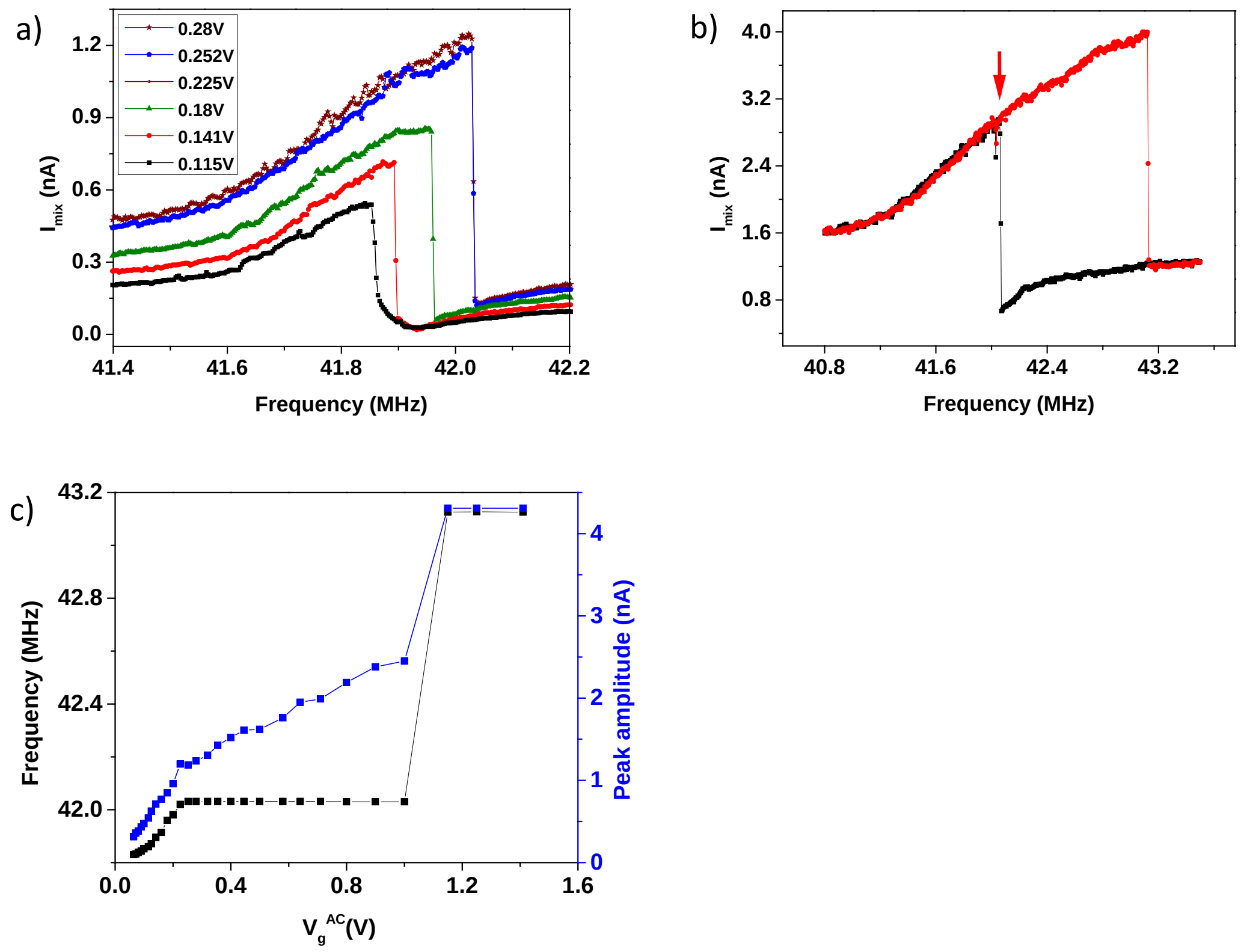

**Figure S7**: *Probing internal resonance using $1\omega$ method. a) Dynamic response of the fundamental mode in nonlinear regime and pinning of frequency due to the first internal resonance. b) Response of the device during forward (red) and backward (black) frequency sweeps. The drive levels were enough to drive the device into second internal resonance. c) Pinning of frequency during the first and second internal resonance. Increase in peak amplitude during the first internal resonance is likely due to the change in the electrical background.*

## Coupling between modes:

In our device to evaluate the strength of the coupling between the fundamental and higher modes, we look at change in resonance frequency of the higher mode while driving the fundamental mode. Modes 4 and 5 show strong frequency dispersion with the amplitude of the first mode. Modes 2, 3 and 6 typically show much smaller frequency dispersion (figure S8).

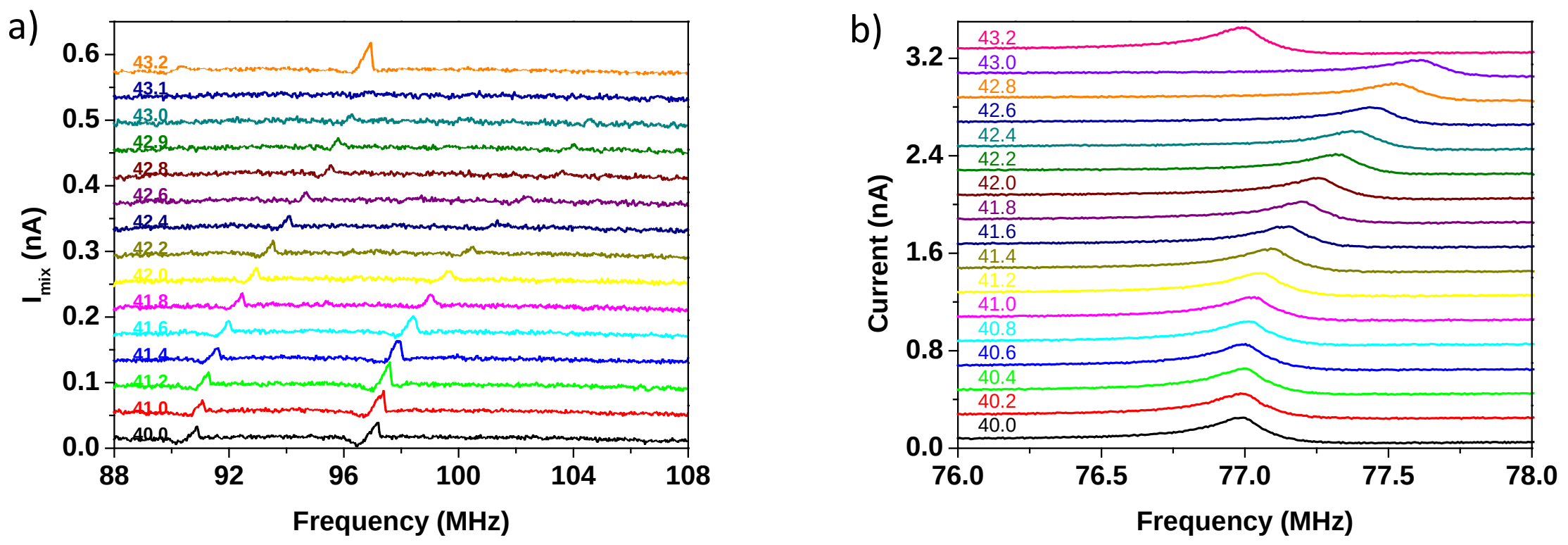

**Figure S8**: *Effect of motion of 1st mode on the response of the higher modes. The drive level of the 1st mode is adjusted to put it in the 2nd internal resonance. a) The resonant frequencies of the mode 4 and 5 are strongly affected by the amplitude of mode 1. Measurements were done using 2$\omega$ technique. b) Other modes that were experimentally probed (mode 2, 3 and 6) showed comparatively smaller effect on the resonant frequency. Measurements were done using 1$\omega$ technique.*

**Splitting of mode 4 due to 1:1 internal resonance:**

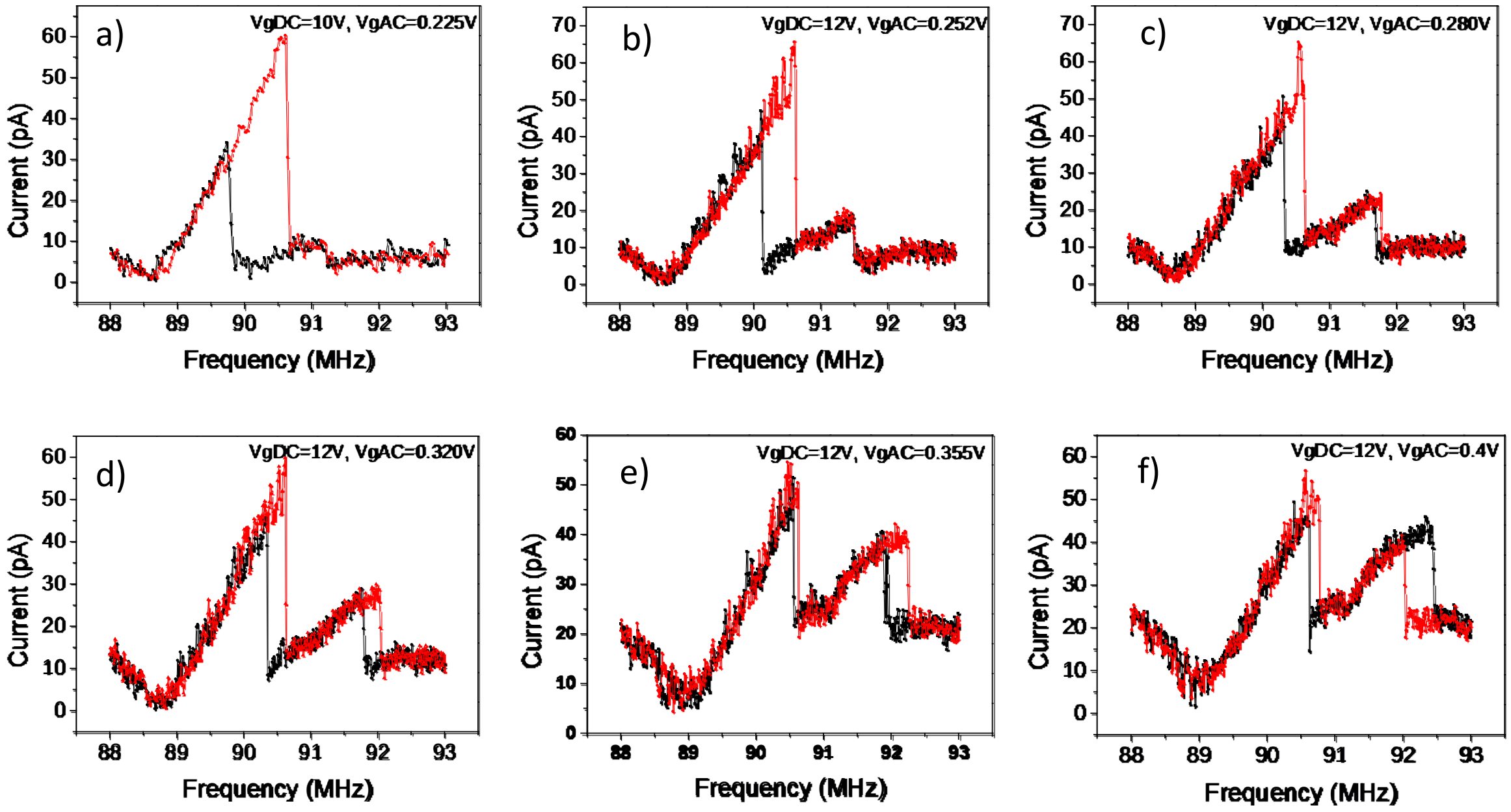

**Figure S9**: *1:1 internal resonance between mode 4 and 5. Response of mode 4 as the drive level is increased.*